\documentclass[aps,prl,reprint,superscriptaddress,nofootinbib,floatfix]{revtex4-2}
\usepackage{amsmath,amssymb}
\usepackage{graphicx}
\usepackage{xcolor}
\usepackage{aas_macros}
\usepackage{threeparttable}
\usepackage{hyperref}
\hypersetup{hidelinks}

\usepackage{hyperref}
\hypersetup{
	colorlinks   = true, 
	urlcolor     = cyan, 
	linkcolor    = red, 
	citecolor   = blue 
}

\newcommand{\changed}[1]{{#1}}

\begin{document}
\title{Evidence for an extended Galactic proton component beyond the cosmic-ray knee}
\author{Jieshuang Wang}
 \email{jieshuangwang@sjtu.edu.cn}
\affiliation{State Key Laboratory of Dark Matter Physics, Tsung-Dao Lee Institute \& School of Physics and Astronomy, Shanghai Jiao Tong University, Shanghai 201210, China}
\begin{abstract}
The cosmic-ray (CR) \textit{knee} at a few petaelectronvolts (PeV) is commonly associated with the acceleration limit of Galactic protons.
Recent detection of multi-PeV photons from the microquasar Cygnus X-3 suggests proton acceleration to tens of PeV.
However, the contribution of such type of sources to CRs remains unclear.
Here we analyse proton and helium spectral data spanning from 0.1 to 1000~PeV and find that the knee strongly disfavours a single terminal cutoff.
Full-spectrum analyses incorporating measurements down to tens of gigaelectronvolts (GeV) require a proton contribution extending beyond $100\,\mathrm{PeV}$.
This contribution either extends continuously from GeV to 100~PeV energies or belongs to a distinct component whose spectrum has a spectral break at the knee.
Within source-population interpretations, its connection to lower-energy Galactic cosmic rays suggests a Galactic origin.
The component from GeV energies is conventionally associated with supernova remnants, but extending it beyond $100\,\mathrm{PeV}$ challenges acceleration theory and $\gamma$-ray observations of ordinary remnants .
A spectral break, by contrast, can arise naturally from rigidity-dependent transport.
This interpretation makes the knee a probe of Galactic CR transport, and connects the extended proton spectrum to extreme Galactic accelerators, such as super-Eddington X-ray binaries, that can also produce PeV $\gamma$ rays and neutrinos. 
\end{abstract}
\maketitle

CRs are cosmic messengers composed predominantly of energetic atomic nuclei, including protons, helium and heavier ions.
Their all-particle spectrum follows an approximate power law, $\Phi(E)\propto E^{-\gamma}$, characterized by several spectral breaks, where $\gamma$ is the spectral index~\cite{Hillas_ARA&A_1984}.
A pronounced steepening at a few PeV, known as the CR \textit{knee}, is a defining feature of Galactic CRs.
In the standard picture, supernova remnants (SNRs) supply the dominant Galactic CR population from GeV energies to the knee with an approximately power-law spectrum shaped by diffusion in interstellar turbulence, and the knee is then usually attributed to the maximum attainable energy in SNRs ~\cite{Blasi_A&ARv_2013,Amato_IJMPD_2014,Gabici_IJMPD_2019},
although a change in the dominant transport mechanism has also been suggested~\cite{Horandel_APh_2004}.
For a common maximum rigidity, $R_{\max}$, nuclei of charge $Ze$ can reach higher energies, $E_{\max}\simeq ZeR_{\max}$, where rigidity is defined as $R=pc/(Ze)$ and $p$ is the particle momentum.
The resulting sequence of elemental cutoffs will lead to an increasingly heavy composition above the proton knee, and the Galactic iron component would extend to around $100~$PeV, near the so-called \textit{second knee} of the all-particle CR spectrum~\cite{KASCADE_PhRvL_2011}, broadly consistent with observations~\cite{KASCADE_PhRvL_2011,Horandel_APh_2004,Gabici_IJMPD_2019}. This correspondence strengthens the conventional interpretation of the knee as the proton acceleration limit of SNRs, the dominant Galactic sources in the standard picture.


Precision measurements have nevertheless revealed substantially richer structure beyond this simple scenario of the dominant Galactic population, especially the proton and helium measurements, which constitute the primary light CR component.
Separating these nuclei reveals charge-dependent features that are blended in the all-particle spectrum.
Space-borne measurements by AMS-02~\cite{AMS_PhRvL_2015a,AMS_PhRvL_2015b,AMS_PhR_2021}, PAMELA~\cite{PAMELA_Sci_2011}, CALET~\cite{CALET_PhRvL_2022,CALET_PhRvL_2023} and DAMPE~\cite{DAMPE_SciA_2019,DAMPE_PhRvL_2021} have established spectral hardenings at hundreds of gigavolts (GV) and softenings at multi-teravolt (TV) rigidities.
The latest DAMPE measurements further reveal a common softening near $15\,\mathrm{TV}$ from protons to irons, demonstrating the charge dependence of this structure~\cite{DampeCollaboration_Natur_2026}.
At higher energies, the Large High Altitude Air Shower Observatory (LHAASO) has measured the proton and helium spectra across their PeV knees with high precision~\cite{Cao_SciBu_2025,Cao_PhRvL_2026}.
The proton spectrum hardens near $0.4\,\mathrm{PeV}$ and then softens above $3.3\,\mathrm{PeV}$~\cite{Cao_SciBu_2025}, and the corresponding features are also present in helium~\cite{Cao_PhRvL_2026}.

Recent $\gamma$-ray observations of microquasars, jetted X-ray binaries (XRBs), provide evidence for powerful Galactic accelerators~\cite{HESS_SS433,HAWC_V4641,LhaasoCollaboration_NSRev_2025,LHAASO_CygX3_arXiv_2025}.
In particular, LHAASO has detected variable and orbit-modulated gamma-rays reaching $3.7$~PeV from Cygnus X-3, a super-Eddington XRB~\cite{LHAASO_CygX3_arXiv_2025}, where a hadronic interpretation is strongly favoured and requires proton acceleration to above tens of $\mathrm{PV}$, indicating that Galactic accelerators can exceed the conventional knee scale.
Whether such super-accreting XRBs contribute appreciably to the CR spectrum near and above the knee remains unresolved.
Composition-resolved measurements offer a way to investigate this possibility.
LHAASO~\cite{Cao_SciBu_2025,Cao_PhRvL_2026}, together with KASCADE-Grande~\cite{KASCADE_APh_2005,KASCADE_PhRvL_2011,KASCADE_ICRC_2015} and IceTop/IceCube~\cite{IceTop_IceCube_PhRvD_2019}, now probes proton and helium spectra over approximately $0.1$~PeV to $1000$~PeV.
These measurements can reveal rigidity-dependent structures  that can be obscured by heavier nuclei in the all-particle spectrum.
We therefore test whether the proton knee is consistent with a single terminal cutoff, or requires a spectral break or overlapping source contributions.
Distinguishing these possibilities constrains how acceleration beyond the knee is reflected in the CR spectrum and how Galactic transport may shape the highest-energy light component.

We first study the nature of the knee in rigidity space using the LHAASO \footnote{The LHAASO data are provided with three hadronic reconstruction models, we here use the compress dataset, and we have tested different choices of reconstruction will not affect our results, see the Appendix for details.} and IceTop/IceCube proton and helium spectra and the KASCADE-Grande light-component spectrum.
We consider three models: a cutoff power law (CPL), a smooth broken power law (BPL) and the sum of two CPL components (2CPL).
A cutoff power law has the form $f_{\rm CPL}(R)\propto R^{-\gamma}\exp(-R/R_c)$, where $R_c$ is the cutoff rigidity.
By contrast, a smooth broken power law with $f_{\rm BPL}(R)\propto R^{-\gamma_1}
\left[1+\left(R/R_b\right)^{(\gamma_2-\gamma_1)/s}\right]^{-s}$
approaches indices $\gamma_1$ and $\gamma_2$ below and above the rigidity break $R_b$, respectively, while $s$ controls the smoothness.
For each model, we maximize the likelihood and compare the likelihood-ratio test statistic (TS) and the Akaike and Bayesian information criteria (AIC and BIC; see the Appendix for details).

\begin{figure*}[t]
\centering
\includegraphics[width=\textwidth]{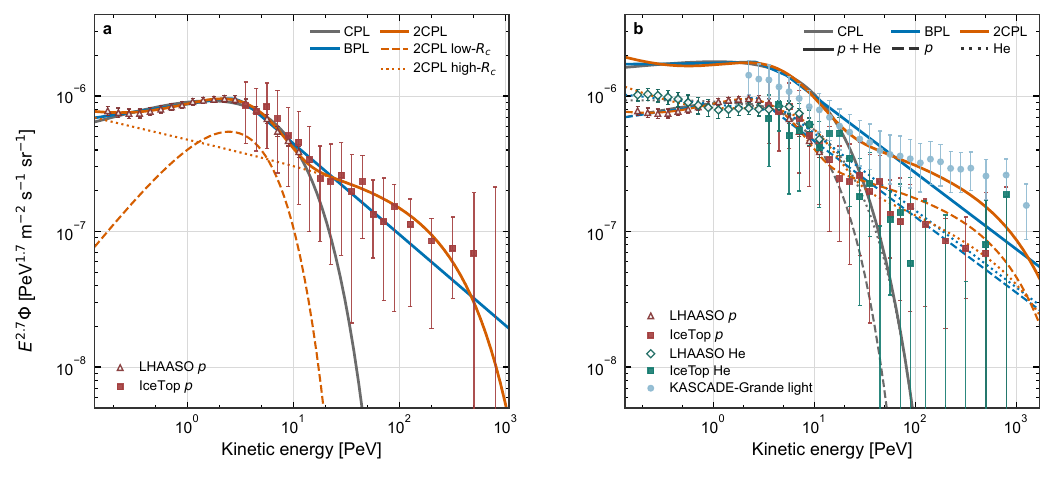}
\caption{\label{fig:knee-morphology}\textbf{Knee morphology with the proton and helium spectra.} (a) Proton-only comparison of the CPL, BPL and 2CPL models using LHAASO and IceTop data. Dashed and dotted curves show the low- and high-rigidity 2CPL components, respectively. \changed{Posterior medians and 16th-to-84th percentile intervals are: CPL $(\gamma_p,R_c)=(2.47^{+0.03}_{-0.03},7.15^{+0.70}_{-0.59}\,\mathrm{PV})$; BPL $(\gamma_{p,1},\Delta\gamma_p,R_b)=(2.58^{+0.03}_{-0.03},0.86^{+0.14}_{-0.10},3.22^{+0.47}_{-0.36}\,\mathrm{PV})$; and 2CPL $(\gamma_{p,1},R_{c,1};\gamma_{p,2},R_{c,2})=(1.76^{+0.25}_{-0.29},2.62^{+0.77}_{-0.59}\,\mathrm{PV};2.88^{+0.04}_{-0.04},306^{+296}_{-142}\,\mathrm{PV})$.} (b) Joint $p$/He knee comparison with a shared cutoff or break rigidity. LHAASO and IceTop provide the composition-resolved $p$ and He measurements, and KASCADE-Grande provides the $p+\mathrm{He}$ light component. The colour distinguishes CPL, BPL and 2CPL, while solid, dashed and dotted curves denote $p+\mathrm{He}$, $p$ and He, respectively. \changed{The shared posterior scales are $R_c=8.86^{+0.98}_{-0.83}\,\mathrm{PV}$ (CPL), $R_b=2.76^{+0.25}_{-0.23}\,\mathrm{PV}$ (BPL), and $(R_{c,1},R_{c,2})=(2.58^{+0.52}_{-0.41},823^{+120}_{-170})\,\mathrm{PV}$ (2CPL). For the joint BPL, $(\gamma_{p,1},\Delta\gamma_p)=(2.59^{+0.02}_{-0.03},0.69^{+0.06}_{-0.06})$ and $(\gamma_{{\rm He},1},\Delta\gamma_{\rm He})=(2.78^{+0.02}_{-0.02},0.51^{+0.08}_{-0.07})$.} Complete constraints and model comparisons are given in Appendix Table~\ref{tab:knee-model-comparison}.}
\end{figure*}

We first test the models with the proton-only data.
Figure~\ref{fig:knee-morphology}a and Appendix Table~\ref{tab:knee-model-comparison} show the proton-only model comparison.
The CPL model gives a cutoff rigidity of $R_c=7.15^{+0.70}_{-0.59}\,\mathrm{PV}$ with $\gamma=2.47\pm0.03$.
Compared with the CPL model, the BPL model improves the likelihood by $\Delta\mathrm{TS}=25.66$ with two additional free parameters, corresponding to a nominal significance of $4.6\sigma$.
The corresponding AIC and BIC decreases are 21.67 and 18.34, respectively, providing strong model-selection evidence against a single cutoff.
The BPL model gives a rigidity break at $R_b=3.22^{+0.47}_{-0.36}\,\mathrm{PV}$ and a steepening $\Delta\gamma=\gamma_2-\gamma_1=0.86^{+0.14}_{-0.10}$.
The 2CPL model instead describes the knee with a narrow component having $\gamma_1=1.76^{+0.25}_{-0.29}$ and $R_{c,1}=2.62^{+0.77}_{-0.59}\,\mathrm{PV}$, while a broader component extends to $R_{c,2}=306^{+296}_{-142}\,\mathrm{PV}$.
This model is also favoured over CPL, with AIC and BIC decreases of 22.50 and 17.51 and $\Delta\mathrm{TS}=28.50$, corresponding to a nominal significance of $4.5\sigma$.

The same morphology test becomes more constraining when the proton and helium data are fitted together.
We fitted the $p$, He and $p+\mathrm{He}$ spectra simultaneously with the same rigidity cutoff or break for corresponding subcomponents, namely $R_{c/b,p}=R_{c/b,\rm He}$.
The proton and helium spectral indices are fitted independently, as motivated by studies of particle acceleration~\cite{Ohira_ApJL_2011,Malkov_PhRvL_2012,Ptuskin_ApJ_2013} and Galactic CR propagation~\cite{Evoli_PhysRevD_2019,Korsmeier_PhysRevD_2022}.
The results are presented in panel b of Figure~\ref{fig:knee-morphology} and Appendix Table~\ref{tab:knee-model-comparison}.
The joint fits recover the same qualitative structures as the proton-only fits, with quantitative shifts in some parameters, and the posterior medians of the proton and helium spectral indices differ by no more than approximately 0.2, supporting a common origin.

With the joint proton and helium data, both the BPL and 2CPL models are strongly favoured over the CPL model, as shown in Appendix Table~\ref{tab:knee-model-comparison}.
For BPL, AIC and BIC decrease by 103.54 and 95.61, respectively; for 2CPL, they decrease by 115.10 and 101.88.
For the BPL model, the joint rigidity break is $R_b=2.76^{+0.25}_{-0.23}\,\mathrm{PV}$, with changes in spectral index below and above the break at $\Delta\gamma=0.51$ for helium and $0.69$ for protons.
For the 2CPL model, a narrow component around the knee is again favoured, with $R_{c,1}=2.58^{+0.52}_{-0.41}\,\mathrm{PV}$ and $\gamma_1$ ranging from $1.77$ to $1.85$ for protons and helium.
The second component is softer, with $\gamma_2$ ranging from $2.88$ to $3.01$, while its cutoff rigidity reaches $R_{c,2}=823^{+120}_{-170}\,\mathrm{PV}$, far beyond the knee.
These comparisons strongly disfavour a single terminal cutoff, while supporting both a rigidity break and overlapping cutoff components as descriptions for the knee.

To determine how the flux above the knee connects to the broader CR spectrum, we extend the analysis down to approximately $45\,\mathrm{GV}$ using AMS-02~\cite{AMS_PhRvL_2015a,AMS_PhRvL_2015b,AMS_PhR_2021} and DAMPE~\cite{DAMPE_SciA_2019,DAMPE_PhRvL_2021,DampeCollaboration_Natur_2026}.
The resulting proton and helium dataset spans kinetic energies from tens of $\mathrm{GeV}$ to approximately $1000\,\mathrm{PeV}$.
We modelled the full spectra using two alternative parameterizations: a sum of three CPL components (3CPL) and a model comprising two CPL components plus a smoothly broken power law (2CPL+BPL).
In each model, one CPL component phenomenologically captures the multi-TV bump defined by the spectral hardening at several hundred GV and the subsequent softening near $15\,\mathrm{TV}$.
Such an additional component may naturally correspond to a nearby source~\cite{Savchenko_ApJL_2015,Ahlers_PhRvL_2016,Liu_JCAP_2019,Yuan_PhRvD_2026} or a different source population~\cite{Recchia_A&A_2024}. Besides, it may also provide an effective description of spectral distortions caused by propagation effects~\cite{Chernyshov_ApJ_2022} or the reacceleration by a nearby star~\cite{Malkov_ApJ_2022}. 
In the latter cases, the fitted CPL parameters for the multi-TV excess should not be interpreted directly as the injection spectrum or maximum rigidity of an independent source population.
Apart from the multi-TV excess, other fitted components may represent different source populations, and such multiple-population descriptions for Galactic CRs have been widely explored, e.g., Refs.~\cite{Vieu_MNRAS_2023,Recchia_A&A_2024,Kaci_arXiv_2025,Zhang_PhRvD_2025,Yuan_PhRvD_2026,Aharonian_arXiv_2026}.

\begin{figure}[t]
\centering
	\includegraphics[width=\columnwidth]{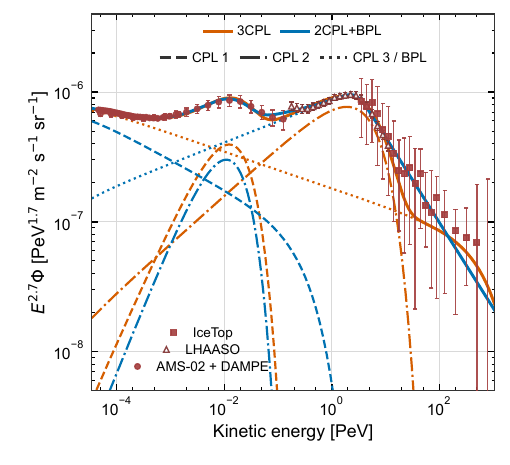}
\caption{\label{fig:proton-full-components}\textbf{Model comparison for the full proton spectrum data.} AMS-02 and DAMPE provide the lower-energy spectrum, while LHAASO and IceTop/IceCube constrain the knee and higher-energy tail. Solid orange and blue curves show the 3CPL and 2CPL+BPL models, respectively; dashed, dash-dotted and dotted curves show their first, second and highest-rigidity components. The two models provide comparably good descriptions despite their different decompositions, but both contain a high-rigidity component that extends beyond the knee. \changed{Posterior medians and 16th-to-84th percentile intervals for the 3CPL knee and high-rigidity components are $(\gamma_{p,\mathrm{knee}},R_{c,\mathrm{knee}})=(2.33^{+0.04}_{-0.05},5.18^{+0.58}_{-0.49}\,\mathrm{PV})$ and $(\gamma_{p,\mathrm{high}},R_{c,3})=(2.84^{+0.02}_{-0.01},602.64^{+252.70}_{-245.89}\,\mathrm{PV})$; for the 2CPL+BPL broken component they are $(\gamma_{p,\mathrm{BPL},1},\Delta\gamma_{p,\mathrm{BPL}},R_{b,\mathrm{BPL}})=(2.56^{+0.03}_{-0.05},0.89^{+0.16}_{-0.11},3.17^{+0.49}_{-0.36}\,\mathrm{PV})$.} Complete constraints and model comparisons are given in Appendix Table~\ref{tab:proton-full-spectrum}.}
\end{figure}

We tested these two scenarios with the full proton dataset. Figure~\ref{fig:proton-full-components} and Appendix Table~\ref{tab:proton-full-spectrum} compare the 3CPL and 2CPL+BPL scenarios and display their individual components.
The two scenarios describe the data comparably well: AIC favours 2CPL+BPL by 2.09, whereas BIC favours the lower-dimensional 3CPL model by 2.87.
Apart from the bump-like feature around $10\,\mathrm{TV}$, both contain a relatively hard contribution around the knee.
In 2CPL+BPL, the knee is a spectral break at $R_{b,\mathrm{BPL}}=3.17^{+0.49}_{-0.36}\,\mathrm{PV}$; the component has a low-energy index $\gamma_{p,\mathrm{BPL},1}=2.56^{+0.03}_{-0.05}$ and steepens by $\Delta\gamma_{p,\mathrm{BPL}}=0.89^{+0.16}_{-0.11}$.
In 3CPL, a component with $\gamma_{p,\mathrm{knee}}=2.33^{+0.04}_{-0.05}$ cuts off at $R_{c,\mathrm{knee}}=5.18^{+0.58}_{-0.49}\,\mathrm{PV}$.
The model also contains a component extending from tens of GV to hundreds of PV, with $\gamma_{p,\mathrm{high}}=2.84^{+0.02}_{-0.01}$ and a cutoff at $R_{c,3}=602.64^{+252.70}_{-245.89}\,\mathrm{PV}$.
Both descriptions therefore require an extended proton contribution beyond \(100\,\mathrm{PV}\), although the proton-only comparison does not decisively select between their different decompositions (Appendix Table~\ref{tab:proton-full-spectrum}).

In the 3CPL scenario, the high-rigidity contribution extends continuously from GeV energies, whereas in the 2CPL+BPL description, a distinct component is present below the knee and steepens across it.
Within these source-population interpretations, the connection to the lower-energy Galactic spectrum suggests a Galactic association for the extended contribution.
At their best-fitting parameters, the extrapolated proton fluxes for both scenarios are below those inferred in phenomenological models of the Pierre Auger spectrum and composition data at EeV energy~\cite{PAO_JCAP_2023,PAO_PRL_2020}, therefore the high-rigidity component above the knee won't overproduce CRs at EeV energies.
	A joint full-spectrum fit to the proton and helium spectra recovered the same extended contribution.
	The 2CPL+BPL description improves the joint likelihood by $\Delta\mathrm{TS}=4.28$, but AIC differs by only 1.72 and BIC favours the lower-dimensional 3CPL model by 11.66 (Appendix Table~\ref{tab:joint-full-spectrum-comparison}).
Overall, the statistical criteria, namely the TS, AIC and BIC, do not consistently favour one scenario over the other for the proton-only and joint $p$/He results, we therefore retain both interpretations and note that both require an extended high-rigidity proton contribution.
These high-rigidity contributions would be difficult to isolate from the all-particle spectrum because the rigidity-ordered contributions of heavier nuclei increasingly dominate above approximately $10\,\mathrm{PeV}$ and can conceal these low-flux tails.

Theoretical considerations and $\gamma$-ray observations further guide the physical interpretation of the two decompositions.
In 3CPL, the high-rigidity component extends the population that dominates the Galactic proton spectrum from GeV-TeV energies (Fig.~\ref{fig:proton-full-components}).
This population is conventionally associated with SNRs.
However, theoretical estimates and $\gamma$-ray observations indicate that ordinary SNR shocks struggle to accelerate protons beyond approximately $10\,\mathrm{PV}$~\cite{Zirakashvili_ApJ_2008,Bell_MNRAS_2013,Blasi_A&ARv_2013,Amato_IJMPD_2014,Gabici_IJMPD_2019,Aharonian_NatAs_2019}.
Extending the same population to $100\,\mathrm{PV}$ is therefore difficult to accommodate within the standard SNR scenario. 
The 3CPL interpretation would require substantially more efficient acceleration by this population, or an additional contribution absorbed into its phenomenological high-rigidity component.

Guided by these considerations, we examine a Galactic transport interpretation of the rigidity-break (2CPL+BPL ) description. This provides a physically motivated explanation of the extended contribution.
If Galactic XRB outflows supply this contribution, their acceleration capability could even exceed the tens-of-PV scale inferred for Cygnus X-3.
For mildly relativistic outflows, the corresponding Hillas-Lovelace power requirement is
\begin{equation}
\label{eq:hillas-lovelace}
L_{\rm K}\gtrsim10^{41}
\left(\frac{R_{\max}}{100\,\mathrm{PV}}\right)^2
\left(\frac{\beta}{0.1}\right)^{-1}
\left(\frac{\sigma_{B}}{0.1}\right)^{-1}
\mathrm{erg\,s^{-1}},
\end{equation}
where $\beta=v/c$ is the dimensionless outflow speed and $\sigma_{B}$ is the magnetization~\cite{Hillas_ARA&A_1984,Lovelace_Natur_1976,Waxman_PhRvL_1995,Blandford_PhST_2000,Wang_ApJL_2025}.
The powerful super-Eddington XRB outflows are plausible candidates with favourable kinetic power, speed and magnetization, which is supported by recent gamma-ray observations \cite{HESS_SS433,HAWC_V4641,LhaasoCollaboration_NSRev_2025,LHAASO_CygX3_arXiv_2025}, although past activities from the super-massive black hole in the Galactic center and extreme SNRs remain possible \cite{Wang_ApJL_2025}.

If transport produces the spectral break, the knee also becomes a probe of Galactic magnetism.
The transition rigidity would depend on the ordered and turbulent components of the Galactic magnetic field, which regulate CR propagation~\cite{Berezinskii_1990,Candia_JHEP_2002,Ptuskin_A&A_1993,Biermann_A&A_1993,Roulet_IJMPA_2004,Horandel_APh_2004,Giacinti_PhRvD_2015}.
A sharp transition from resonant diffusion, $D_{\rm res}\propto R^{2-q}$, to the non-resonant asymptote, $D_{\rm nonres}\propto R^2$, would produce $\Delta\gamma\simeq q=5/3$ for the Kolmogorov turbulence, substantially larger than observed.
However, diffusion-drift models instead predict transverse diffusion $D_\perp \propto R^{2-q}$ below the knee and Hall drift $D_A\propto R$ above it.
For an injected spectrum $Q(R)\propto R^{-\alpha}$, the observed indices are $\gamma_1=\alpha+2-q$ and $\gamma_2=\alpha+1$, giving $\Delta\gamma=q-1$.
Kolmogorov turbulence predicts $\Delta\gamma=2/3$, slightly below the proton-only constraint $\Delta\gamma\approx 0.89$.
Interpreted within this model, the proton-only spectrum implies a turbulence index $q\approx1.89$ and an injection index $\alpha\approx2.44$.
We noted that other mechanisms may also be able to produce an effective spectral break~\cite{Horandel_APh_2004}. 
For example, 
a distribution of outflow powers can imply a distribution of attainable rigidities through the Hillas-Lovelace scaling (Eq.~\ref{eq:hillas-lovelace}).

Under such a Galactic interpretation, the extended light component connects composition-resolved CR measurements to multi-messenger studies of the Milky Way.
If XRB outflows contribute above the knee, hadronic interactions in their jets, remnants or surroundings should produce PeV $\gamma$ rays and neutrinos~\cite{HAWC_V4641,LhaasoCollaboration_NSRev_2025,LHAASO_CygX3_arXiv_2025,Peretti_A&A_2025,Ohira_MNRAS_2025,Wang_ApJL_2025,Zhang_arXiv_2026,Bykov_PhRvD_2025,Wei_arXiv_2025}.
Variability and/or morphology could identify the acceleration sites, whereas extended emission might trace escaped particles.
If this source population supplies the extended contribution, its maximum rigidity and flux can constrain the high-luminosity tail and duty cycle of Galactic XRBs.
These quantities are controlled by binary mass-transfer histories, the duration of super-Eddington accretion and the efficiency with which accretion power and magnetic flux launch non-thermal jets, linking the CR spectrum to binary evolution and accretion physics~\cite{Mineo_MNRAS_2012,King_NewAR_2023,Wang_ApJL_2025}.


\begin{acknowledgments}
We acknowledge the use of OpenAI's ChatGPT 5.2 and 6 Astra for assistance with code development and debugging and for language editing. All AI-assisted code and outputs were reviewed and validated by the authors.
\end{acknowledgments}

\bibliographystyle{apsrev4-2}
\bibliography{main}
\appendix
\clearpage
\onecolumngrid

\section{Supplemental Materials }
\subsubsection{Model parameterization}
We describe the spectra in rigidity space so that charge-dependent structure can be compared directly between protons, helium and the light component.
For charge number $Z$ and rest energy $m$, energies expressed in $\mathrm{PeV}$ and rigidity in $\mathrm{PV}$ obey $E_{\rm tot}=E_k+m$ and $R=\sqrt{E_{\rm tot}^2-m^2}/Z$.
The differential flux is therefore
\begin{equation}
\Phi(E_k)=f(R)\,\frac{dR}{dE_k}=f(R)\,\frac{E_{\rm tot}}{Z^2 R}.
\end{equation}
We normalize the rigidity dependence at $R_0=0.1\,\mathrm{PV}$.
The normalization $F$ has units of differential rigidity flux, $\mathrm{PV}^{-1}\,\mathrm{m}^{-2}\,\mathrm{s}^{-1}\,\mathrm{sr}^{-1}$.
In the fits it is sampled through the dimensionless parameter $\log_{10}(F/F_0)$, where $F_0=1\,\mathrm{PV}^{-1}\,\mathrm{m}^{-2}\,\mathrm{s}^{-1}\,\mathrm{sr}^{-1}$.
The cutoff power-law building block is
\begin{equation}
f_{\rm CPL}(R;F,\gamma,R_c)=F\left(\frac{R}{R_0}\right)^{-\gamma}\exp\left(-\frac{R}{R_c}\right),
\end{equation}
and the smooth broken power law is
\begin{equation}
\begin{aligned}
f_{\rm BPL}(R;F,\gamma_{1},\gamma_{2},R_b,s)
&=F\left(\frac{R}{R_0}\right)^{-\gamma_{1}} 
\times\left[1+\left(\frac{R}{R_b}\right)^{(\gamma_{2}-\gamma_{1})/s}\right]^{-s}.
\end{aligned}
\end{equation}
The knee-only 2CPL model is $f(R)=\sum_{i=1}^{2}f_{\rm CPL}(R;F_i,\gamma_i,R_{c,i})$.
For the full-spectrum fits, 3CPL is written as
\begin{equation}
f_{\rm 3CPL}(R)=\sum_{i=1}^{3}f_{\rm CPL}(R;F_i,\gamma_i,R_{c,i}).
\end{equation}
The hybrid model is
\begin{equation}
\begin{aligned}
f_{\rm 2CPL+BPL}(R)
&=\sum_{i=1}^{2}f_{\rm CPL}(R;F_i,\gamma_i,R_{c,i}) 
+f_{\rm BPL}(R;F_3,\gamma_{3,1},\gamma_{3,2},R_{b,3},s_3).
\end{aligned}
\end{equation}

\begin{table*}
	\centering
	\caption{Parameter constraints and model comparison for the proton-only data and joint $p$/He data. Here $\log\equiv\log_{10}$, $\log R_c\equiv\log_{10}(R_c/\mathrm{PV})$ and $F_0=1\,\mathrm{PV}^{-1}\,\mathrm{m}^{-2}\,\mathrm{s}^{-1}\,\mathrm{sr}^{-1}$, so each fitted logarithmic normalization is the dimensionless quantity $\log(F/F_0)$. Priors are uniform over the stated intervals. Parameter values are posterior medians with 16th-to-84th percentile intervals. The 2CPL components are ordered by cutoff rigidity, so Component~2 has the higher cutoff. The $\Delta\mathrm{TS}$ values and nominal significances are relative to CPL.}
	\label{tab:knee-model-comparison}
	\scriptsize
	\renewcommand{\arraystretch}{1.03}
	\begin{tabular*}{\textwidth}{@{\extracolsep{\fill}}lllcc}
		\hline
		Model & Parameter & Prior & Posterior ($p$) & Posterior (joint) \\
		\hline
		CPL & $\log(F_p/F_0)$ & $(-5,-1)$ & $-3.51^{+0.03}_{-0.03}$ & $-3.47^{+0.02}_{-0.02}$ \\
		& $\gamma_p$ & $(1,4)$ & $2.47^{+0.03}_{-0.03}$ & $2.53^{+0.03}_{-0.03}$ \\
		& $\log(F_{\rm He}/F_0)$ & $(-5,-1)$ & $\ldots$ & $-3.84^{+0.02}_{-0.02}$ \\
		& $\gamma_{\rm He}$ & $(1,4)$ & $\ldots$ & $2.69^{+0.03}_{-0.02}$ \\
		& $R_c$ [$\mathrm{PV}$] & $\log R_c\in(0,2)$ & $7.15^{+0.70}_{-0.59}$ & $8.86^{+0.98}_{-0.83}$ \\
		\hline
		BPL & $\log(F_p/F_0)$ & $(-5,-1)$ & $-3.48^{+0.03}_{-0.03}$ & $-3.47^{+0.02}_{-0.02}$ \\
		& $\gamma_{p,1}$ & $(2,4)$ & $2.58^{+0.03}_{-0.03}$ & $2.59^{+0.02}_{-0.03}$ \\
		& $\Delta\gamma_p=\gamma_{p,2}-\gamma_{p,1}$ & $\gamma_{p,2}\in(2,5)$ & $0.86^{+0.14}_{-0.10}$ & $0.69^{+0.06}_{-0.06}$ \\
		& $\log(F_{\rm He}/F_0)$ & $(-5,-1)$ & $\ldots$ & $-3.82^{+0.02}_{-0.02}$ \\
		& $\gamma_{{\rm He},1}$ & $(2,4)$ & $\ldots$ & $2.78^{+0.02}_{-0.02}$ \\
		& $\Delta\gamma_{\rm He}=\gamma_{{\rm He},2}-\gamma_{\rm He,1}$ & $\gamma_{{\rm He},2}\in(2,5)$ & $\ldots$ & $0.51^{+0.08}_{-0.07}$ \\
		& $R_b$ [$\mathrm{PV}$] & $(0.3,20)$ & $3.22^{+0.47}_{-0.36}$ & $2.76^{+0.25}_{-0.23}$ \\
		& $s$ & $(0.01,1)$ & $0.15^{+0.15}_{-0.09}$ & $0.05^{+0.05}_{-0.03}$ \\
		\hline
		2CPL & $\log(F_{p,1}/F_0)$ & $(-6,-1)$ & $-4.45^{+0.30}_{-0.32}$ & $-4.44^{+0.23}_{-0.24}$ \\
		& $\gamma_{p,1}$ & $(1,4)$ & $1.76^{+0.25}_{-0.29}$ & $1.77^{+0.20}_{-0.21}$ \\
		& $\log(F_{{\rm He},1}/F_0)$ & $(-6,-1)$ & $\ldots$ & $-4.97^{+0.23}_{-0.25}$ \\
		& $\gamma_{{\rm He},1}$ & $(1,4)$ & $\ldots$ & $1.85^{+0.20}_{-0.22}$ \\
		& $R_{c,1}$ [$\mathrm{PV}$] & $\log R_{c,1}\in(0,1)$ & $2.62^{+0.77}_{-0.59}$ & $2.58^{+0.52}_{-0.41}$ \\
		& $\log(F_{p,2}/F_0)$ & $(-6,-1)$ & $-3.45^{+0.05}_{-0.09}$ & $-3.45^{+0.05}_{-0.07}$ \\
		& $\gamma_{p,2}$ & $(1,4)$ & $2.88^{+0.04}_{-0.04}$ & $2.88^{+0.02}_{-0.03}$ \\
		& $\log(F_{{\rm He},2}/F_0)$ & $(-6,-1)$ & $\ldots$ & $-3.82^{+0.03}_{-0.04}$ \\
		& $\gamma_{{\rm He},2}$ & $(1,4)$ & $\ldots$ & $3.01^{+0.03}_{-0.03}$ \\
		& $R_{c,2}$ [$\mathrm{PV}$] & $\log R_{c,2}\in(\log R_{c,1},3)$ & $306^{+296}_{-142}$ & $823^{+120}_{-170}$ \\
		\hline
	\end{tabular*}
	
	\vspace{0.6em}
	\begin{tabular*}{\textwidth}{@{\extracolsep{\fill}}llcccccc}
		\hline
		Dataset & Model & $k$ & $\ln\mathcal{L}_{\max}$ & $\Delta\mathrm{TS}$ & AIC & BIC & Nominal $\sigma$ \\
		\hline
		$p$-only & CPL & 3 & $-15.50$ & $\ldots$ & $37.01$ & $42.00$ & $\ldots$ \\
		& BPL & 5 & $-2.67$ & $25.66$ & $15.34$ & $23.66$ & $4.6\,\sigma$ \\
		& 2CPL & 6 & $-1.26$ & $28.50$ & $14.51$ & $24.49$ & $4.5\,\sigma$ \\
		\hline
		Joint $p$/He & CPL & 5 & $-81.87$ & $\ldots$ & $173.74$ & $186.97$ & $\ldots$ \\
		& BPL & 8 & $-27.10$ & $109.54$ & $70.20$ & $91.36$ & $9.9\,\sigma$ \\
		& 2CPL & 10 & $-19.32$ & $125.10$ & $58.64$ & $85.09$ & $10.3\,\sigma$ \\
		\hline
	\end{tabular*}
\end{table*}

\begin{table*}
	\centering
	\caption{Full proton shape-parameter constraints and model comparison. Parameter values are posterior medians with 16th-to-84th percentile intervals. Cutoffs below $0.1\,\mathrm{PV}$ are reported in $\mathrm{TV}$. Component normalizations are nuisance parameters and are omitted. Model-comparison differences are defined relative to 3CPL. For the 2CPL+BPL model, $\gamma_{\mathrm{BPL},2}=\gamma_{\mathrm{BPL},1}+\Delta\gamma_{\mathrm{BPL}}$. \changed{The high-rigidity cutoff is bounded above by the $1000\,\mathrm{PV}$ prior and should not be read as a precise measurement of the acceleration limit.}}
	\label{tab:proton-full-spectrum}
	\scriptsize
	\renewcommand{\arraystretch}{1.04}
	\begin{tabular*}{\textwidth}{@{\extracolsep{\fill}}lllc}
		\hline
		Model & Parameter & Prior & Posterior  \\
		\hline
		3CPL & $\gamma_{p,10\mathrm{TV}}$ & $(1,4)$ & $1.74^{+0.16}_{-0.18}$ \\
		& $R_{c,10\mathrm{TV}}$ [$\mathrm{TV}$] & $\log_{10}(R_c/\mathrm{PV})\in(-2.5,-0.5)$ & $12.63^{+4.24}_{-3.23}$ \\
		& $\gamma_{p,\mathrm{knee}}$ & $(1,4)$ & $2.33^{+0.04}_{-0.05}$ \\
		& $R_{c,\mathrm{knee}}$ [$\mathrm{PV}$] & $\log_{10}(R_c/\mathrm{PV})\in(0,1)$ & $5.18^{+0.58}_{-0.49}$ \\
		& $\gamma_{p,\mathrm{high}}$ & $(1,4)$ & $2.84^{+0.02}_{-0.01}$ \\
		& $R_{c,3}$ [$\mathrm{PV}$] & $\log_{10}(R_c/\mathrm{PV})\in(-1,3)$ & $602.64^{+252.70}_{-245.89}$ \\
		\hline
		2CPL+BPL & $\gamma_{p,\mathrm{CPL1}}$ & $(2,4)$ & $2.95^{+0.05}_{-0.05}$ \\
		& $R_{c,\mathrm{CPL1}}$ [$\mathrm{PV}$] & $\log_{10}(R_c/\mathrm{PV})\in(-3,0)$ & $0.17^{+0.41}_{-0.15}$ \\
		& $\gamma_{p,\mathrm{CPL2}}$ & $(1,3)$ & $1.66^{+0.27}_{-0.31}$ \\
		& $R_{c,\mathrm{CPL2}}$ [$\mathrm{TV}$] & $\log_{10}(R_c/\mathrm{PV})\in(-3,-1)$ & $11.00^{+5.50}_{-3.60}$ \\
		& $\gamma_{p,\mathrm{BPL},1}$ & $(2,4)$ & $2.56^{+0.03}_{-0.05}$ \\
		& $\Delta\gamma_{p,\mathrm{BPL}}$ & $\gamma_{p,\mathrm{BPL},2}\in(3,5)$ & $0.89^{+0.16}_{-0.11}$ \\
		& $R_{b,\mathrm{BPL}}$ [$\mathrm{PV}$] & $(1,10)$ & $3.17^{+0.49}_{-0.36}$ \\
		& $s_{\mathrm{BPL}}$ & $(0.01,1)$ & $0.17^{+0.17}_{-0.10}$ \\
		\hline
	\end{tabular*}
	
	\vspace{0.6em}
	\begin{tabular*}{\textwidth}{@{\extracolsep{\fill}}lrrrrrrr}
		\hline
		Model & $k$ & $\ln\mathcal{L}_{\max}$ & $\Delta\mathrm{TS}$ & AIC & $\Delta$AIC & BIC & $\Delta$BIC \\
		\hline
		3CPL & 9 & $-9.31$ & $-$ & $36.62$ & $-$ & $58.92$ & $-$ \\
		2CPL+BPL & 11 & $-6.27$ & $6.09$ & $34.54$ & $-2.09$ & $61.79$ & $+2.87$ \\
		\hline
	\end{tabular*}
\end{table*}

\subsubsection{Data sets and likelihood tests}
Rigidity-space measurements are transformed into kinetic energy in $\mathrm{PeV}$, and all fluxes are expressed as $\Phi(E_k)$ in $\mathrm{PeV}^{-1}\,\mathrm{m}^{-2}\,\mathrm{s}^{-1}\,\mathrm{sr}^{-1}$.
The low-energy baselines are constructed from AMS-02 and DAMPE~\cite{AMS_PhRvL_2015a,AMS_PhRvL_2015b,AMS_PhR_2021,DAMPE_SciA_2019,DAMPE_PhRvL_2021,DampeCollaboration_Natur_2026}; the knee region is constrained by LHAASO and IceTop~\cite{IceTop_IceCube_PhRvD_2019,Cao_SciBu_2025,Cao_PhRvL_2026}; and the high-energy light component is taken from KASCADE-Grande~\cite{KASCADE_PhRvL_2011,KASCADE_ICRC_2015}.
We do not recalibrate these published data or adjust their relative normalizations.

For a model prediction $\Phi_{\rm mod}(E_i|\boldsymbol{\theta})$ and a measured flux $\Phi_i$, we define the residual $\Delta_i=\Phi_{\rm mod}(E_i|\boldsymbol{\theta})-\Phi_i$.
The likelihood is defined as
\begin{equation}
\ln\mathcal{L}(\boldsymbol{\theta})=-\frac{1}{2}\sum_i\left(\frac{\Delta_i}{\sigma_i}\right)^2.
\end{equation}
For asymmetric measurement errors, we select $\sigma_i=\sigma_{i,+}$ when $\Delta_i>0$, corresponding to a model prediction above the measured flux, and $\sigma_i=\sigma_{i,-}$ when $\Delta_i<0$.
This likelihood treats the flux errors as independent and does not include correlations between energy bins or reconstructed species.
\changed{Parameter estimates in the text, figure captions and tables are posterior medians with 16th-to-84th percentile credible intervals. The maximum likelihood is used separately to compute TS, AIC and BIC.}
We sample parameter posteriors using the Python package \texttt{emcee}~\cite{ForemanMackey_PASP_2013}.
We define the likelihood-ratio improvement as
\begin{equation}
\Delta\mathrm{TS}=2\left(\ln\mathcal{L}_{\rm alt}-\ln\mathcal{L}_{\rm null}\right).
\end{equation}
Model-complexity penalties are reported separately through $\Delta\mathrm{AIC}=2\Delta k-\Delta\mathrm{TS}$ and $\Delta\mathrm{BIC}=\Delta k\ln N-\Delta\mathrm{TS}$, where $k$ is the number of free parameters and $N$ the number of fitted data points.
Information-criterion differences are alternative minus reference model, so negative values favour the alternative, whereas positive $\Delta\mathrm{TS}$ denotes a likelihood improvement.
For the knee comparisons, we quote nominal Gaussian-equivalent significances by mapping a $\chi^2_{\Delta k}$ upper-tail probability to a one-sided Gaussian probability.

The LHAASO spectra are provided as QGSJET, EPOS and SIBYLL reconstructions.
We compress the three reconstructions using their mean central value and a total uncertainty $\sigma_{\rm tot}^2=\sigma_{\rm exp}^2+\sigma_{\rm had}^2$, where $\sigma_{\rm exp}$ is the reported uncertainty and $\sigma_{\rm had}$ is the standard deviation of the three central values.
\changed{The results for the spectral data around the knee and the full spectrum data are presented in Tables \ref{tab:knee-model-comparison} and \ref{tab:proton-full-spectrum}.}
To test whether the treatment of the LHAASO data affects our conclusions, we repeated the BPL-CPL comparison using the compressed, QGSJET, EPOS and SIBYLL data sets separately.
Table~\ref{tab:knee-robustness} summarizes these tests.
The resulting $\Delta\mathrm{TS}$ values are 26-43, while the fitted break remains near $R_b\simeq3\,\mathrm{PV}$ and the fitted steepening near $\Delta\gamma\simeq0.8$.
The improvement of BPL over a single CPL is therefore insensitive to these treatments of the public LHAASO data.

\begin{table}
\centering
\caption{Proton-knee robustness check for different choices of LHAASO datasets. $\Delta\mathrm{AIC}$ and $\Delta\mathrm{BIC}$ are BPL model value minus CPL model value.}
\label{tab:knee-robustness}
\scriptsize
\begin{tabular}{lccccc}
\hline
Test & $\Delta\mathrm{TS}$ & $\Delta\mathrm{AIC}$ & $\Delta\mathrm{BIC}$ & $R_b$ [$\mathrm{PV}$] & $\Delta\gamma$ \\
\hline
Compressed & $26$ & $-22$ & $-18$ & $3.1$ & $0.80$ \\
QGSJET only & $35$ & $-31$ & $-28$ & $3.3$ & $0.81$ \\
EPOS only & $43$ & $-39$ & $-36$ & $3.3$ & $0.85$ \\
SIBYLL only & $40$ & $-36$ & $-33$ & $3.1$ & $0.82$ \\
\hline
\end{tabular}
\end{table}

\subsubsection{Full-spectrum consistency tests for the joint proton and helium data}
\begin{figure*}
\centering
\includegraphics[width=\textwidth]{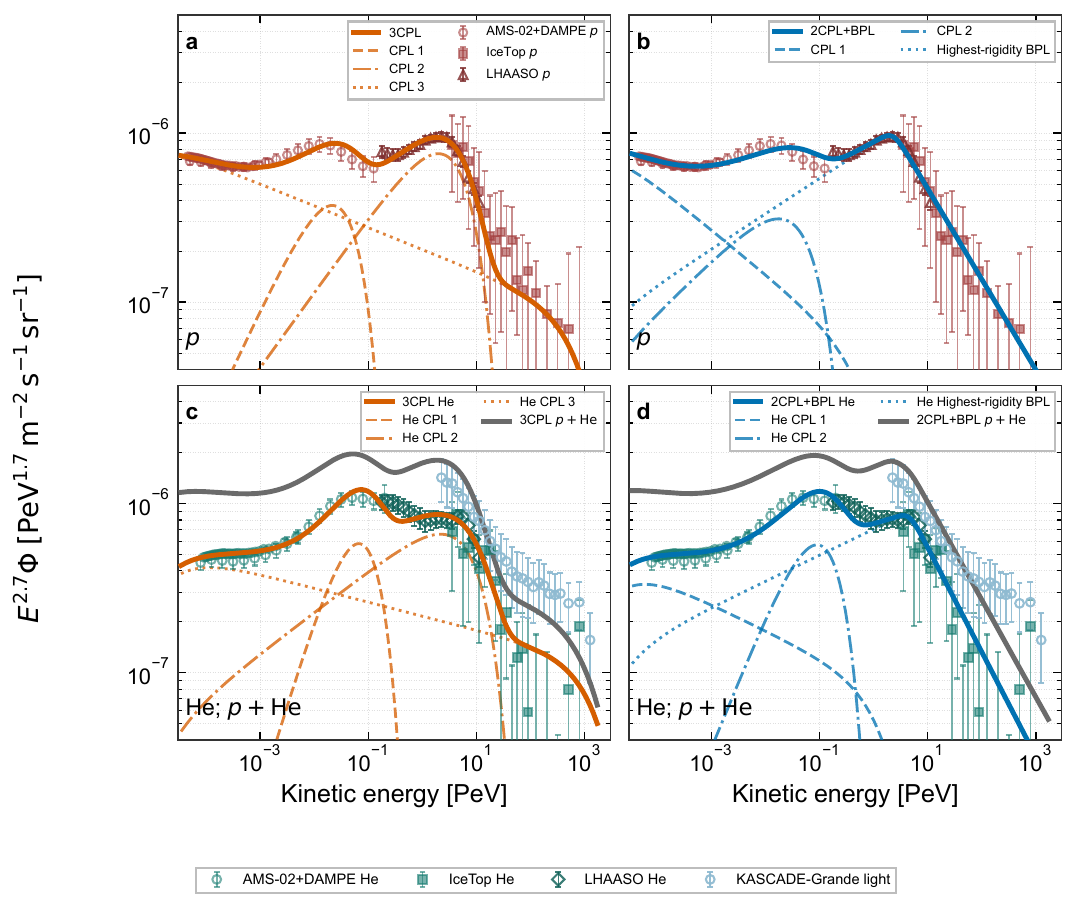}
\caption{\label{fig:joint-full-spectrum-comparison}Full $p$/He/light spectrum including AMS-02 and DAMPE. Curves compare the three-CPL baseline with the hybrid two-CPL plus BPL model. KASCADE-Grande provides the light-component constraint; the LHAASO points show protons and helium separately, and the LHAASO light spectrum is excluded.}
\end{figure*}

As a cross-species consistency check, we repeated the full-spectrum comparison using 203 measurements of $p$, He and $p+\mathrm{He}$.
AMS-02 and DAMPE anchor the lower-energy proton and helium spectra, LHAASO and IceTop constrain the composition-resolved knee region, and KASCADE-Grande supplies the high-energy light-component measurements.
Both models share their rigidity scales between protons and helium but fit the species normalizations and spectral indices independently, without a cross-species similarity prior.
Figure~\ref{fig:joint-full-spectrum-comparison} shows the two model decompositions for the three fitted spectra.
The joint 3CPL and 2CPL+BPL fits reached $\ln\mathcal{L}_{\max}=-50.01$ and $-47.87$, respectively (Table~\ref{tab:joint-full-spectrum-comparison}).
At the likelihood level, 2CPL+BPL improves the fit by $\Delta\mathrm{TS}=4.28$.
AIC is nearly indifferent, with a difference of 1.72 in favour of 3CPL, whereas BIC favours the lower-dimensional 3CPL model by 11.66.
Thus BIC prefers 3CPL for the joint data, while the AIC difference is small; neither criterion identifies the physical sources represented by the fitted components.
We next compared the joint posterior directly with the proton-only constraints to determine which inferred components are stable when helium and light-component data are added.
For 3CPL, the proton index of the highest-rigidity component is unchanged, from $2.84^{+0.02}_{-0.01}$ to $2.83^{+0.01}_{-0.01}$, while the median cutoff moves from $602.64^{+252.70}_{-245.89}\,\mathrm{PV}$ to $881.98^{+81.44}_{-124.60}\,\mathrm{PV}$ within the adopted $1000\,\mathrm{PV}$ analysis range.
The knee-component cutoff shifts from $5.18^{+0.58}_{-0.49}\,\mathrm{PV}$ to $4.41^{+0.31}_{-0.29}\,\mathrm{PV}$, whereas the multi-TV cutoff changes more strongly, from $12.63^{+4.24}_{-3.23}\,\mathrm{TV}$ to $29.65^{+8.69}_{-6.96}\,\mathrm{TV}$.
For 2CPL+BPL, the proton BPL slopes and smoothness remain broadly comparable, although the break rigidity shifts.
Its low-rigidity index changes from $2.56^{+0.03}_{-0.05}$ to $2.49^{+0.04}_{-0.05}$, and its steepening changes from $0.89^{+0.16}_{-0.11}$ to $0.75^{+0.08}_{-0.06}$.
The smoothness changes from $0.17^{+0.17}_{-0.10}$ to $0.16^{+0.10}_{-0.08}$, while the shared break shifts from $3.17^{+0.49}_{-0.36}\,\mathrm{PV}$ to $2.28^{+0.20}_{-0.19}\,\mathrm{PV}$.
By contrast, the lower-rigidity CPL allocation changes substantially, including a shift of its multi-TV cutoff from $11.00^{+5.50}_{-3.60}\,\mathrm{TV}$ to $58.03^{+19.31}_{-14.80}\,\mathrm{TV}$.
Thus the joint fit reproduces the proton-only requirement for an extended high-rigidity contribution and broadly similar high-energy slopes, but not a unique decomposition of the lower-energy components.
The stable result is the extended proton contribution, whereas the detailed component assignments depend on the fitted species and adopted priors.
In particular, the proton-only hybrid fit uses a uniform prior on the high-energy BPL index, while the joint fit uses a uniform prior on the steepening (Tables~\ref{tab:proton-full-spectrum} and \ref{tab:joint-full-spectrum-comparison}).
The comparison therefore tests the persistence of the spectral interpretation across the adopted analyses, rather than isolating the effect of adding helium under identical priors.

\begin{table*}
\centering
\caption{Joint $p$/He/light full-spectrum shape-parameter constraints and model comparison. Posterior values are medians with 16th-to-84th percentile intervals. Here $\log R_c\equiv\log_{10}(R_c/\mathrm{PV})$; cutoffs below $0.1\,\mathrm{PV}$ are reported in $\mathrm{TV}$. Component normalizations are nuisance parameters and are omitted. Model-comparison differences are defined relative to 3CPL.}
\label{tab:joint-full-spectrum-comparison}
\scriptsize
\renewcommand{\arraystretch}{1.02}
\begin{tabular*}{\textwidth}{@{\extracolsep{\fill}}llcc}
\hline
Model & Parameter & Prior & Posterior \\
\hline
3CPL & $\gamma_{p,10\mathrm{TV}}$ & $(1,4)$ & $1.99^{+0.09}_{-0.11}$ \\
 & $\gamma_{\mathrm{He},10\mathrm{TV}}$ & $(1,4)$ & $1.60^{+0.16}_{-0.18}$ \\
 & $R_{c,10\mathrm{TV}}$ [$\mathrm{TV}$] & $\log R_c\in(-2.5,-0.5)$ & $29.65^{+8.69}_{-6.96}$ \\
 & $\gamma_{p,\mathrm{knee}}$ & $(1,4)$ & $2.26^{+0.03}_{-0.04}$ \\
 & $\gamma_{\mathrm{He},\mathrm{knee}}$ & $(1,4)$ & $2.46^{+0.02}_{-0.02}$ \\
 & $R_{c,\mathrm{knee}}$ [$\mathrm{PV}$] & $\log R_c\in(0,1)$ & $4.41^{+0.31}_{-0.29}$ \\
 & $\gamma_{p,\mathrm{high}}$ & $(1,4)$ & $2.83^{+0.01}_{-0.01}$ \\
 & $\gamma_{\mathrm{He},\mathrm{high}}$ & $(1,4)$ & $2.78^{+0.01}_{-0.01}$ \\
 & $R_{c,3}$ [$\mathrm{PV}$] & $\log R_c\in(-1,3)$ & $881.98^{+81.44}_{-124.60}$ \\
\hline
2CPL+BPL & $\gamma_{p,\mathrm{CPL1}}$ & $(2,4)$ & $2.96^{+0.05}_{-0.04}$ \\
 & $\gamma_{\mathrm{He},\mathrm{CPL1}}$ & $(2,4)$ & $2.86^{+0.03}_{-0.04}$ \\
 & $R_{c,\mathrm{CPL1}}$ [$\mathrm{PV}$] & $\log R_c\in(-3,0)$ & $0.57^{+0.29}_{-0.35}$ \\
 & $\gamma_{p,\mathrm{CPL2}}$ & $(1,3)$ & $2.39^{+0.10}_{-0.11}$ \\
 & $\gamma_{\mathrm{He},\mathrm{CPL2}}$ & $(1,3)$ & $1.90^{+0.13}_{-0.15}$ \\
 & $R_{c,\mathrm{CPL2}}$ [$\mathrm{TV}$] & $\log R_c\in(-3,-1)$ & $58.03^{+19.31}_{-14.80}$ \\
 & $\gamma_{p,\mathrm{BPL},1}$ & $(2,4)$ & $2.49^{+0.04}_{-0.05}$ \\
 & $\Delta\gamma_{p,\mathrm{BPL}}$ & $(0.5,2)$ & $0.75^{+0.08}_{-0.06}$ \\
 & $\gamma_{\mathrm{He},\mathrm{BPL},1}$ & $(2,4)$ & $2.54^{+0.03}_{-0.05}$ \\
 & $\Delta\gamma_{\mathrm{He},\mathrm{BPL}}$ & $(0.5,2)$ & $0.80^{+0.08}_{-0.07}$ \\
 & $R_{b,\mathrm{BPL}}$ [$\mathrm{PV}$] & $(1,10)$ & $2.28^{+0.20}_{-0.19}$ \\
 & $s_{\mathrm{BPL}}$ & $(0.01,1)$ & $0.16^{+0.10}_{-0.08}$ \\
\hline
\end{tabular*}

\vspace{0.6em}
\begin{tabular*}{\textwidth}{@{\extracolsep{\fill}}lrrrrrrr}
\hline
Model & $k$ & $\ln\mathcal{L}_{\max}$ & $\Delta\mathrm{TS}$ & AIC & $\Delta$AIC & BIC & $\Delta$BIC \\
\hline
3CPL & 15 & $-50.01$ & $0.00$ & $130.02$ & $0.00$ & $179.71$ & $0.00$ \\
2CPL+BPL & 18 & $-47.87$ & $4.28$ & $131.74$ & $+1.72$ & $191.38$ & $+11.66$ \\
\hline
\end{tabular*}
\end{table*}

\subsubsection{Galactic transport interpretations}
Resonant diffusion gives $D_{\rm res}\propto R^{2-q}$, whereas non-resonant diffusion approaches $D_{\rm nonres}\propto R^2$ at high rigidity.
If the source spectrum is unchanged, a sharp transition predicts
\begin{equation}
\Delta\gamma\simeq 2-(2-q)=q,
\end{equation}
which is about 1.5-1.7 for standard Kraichnan or Kolmogorov turbulence and is larger than observed.
Spatial variations in magnetic coherence length, field geometry and disc-halo transport may broaden a local transition, but the present spectra cannot recover those variations uniquely; we therefore use the diffusion-drift relations only to interpret the measured indices.

The diffusion tensor contains a symmetric transverse coefficient $D_\perp$ and an antisymmetric Hall coefficient $D_A$ \cite{Candia_JHEP_2002,Roulet_IJMPA_2004,Horandel_APh_2004}.
In the diffusion-drift scenario, $D_\perp\propto R^{2-q}$ dominates escape below the knee, whereas $D_A\propto R$ dominates above it.
Here $P(\kappa)\propto\kappa^{-q}$ is the turbulence spectrum and $\kappa$ denotes wavenumber~\cite{Candia_JHEP_2002,Roulet_IJMPA_2004}.
For a source spectrum $Q(R)\propto R^{-\alpha}$, the steady-state spectrum consequently changes from
\begin{equation}
\gamma_1=\alpha+2-q
\end{equation}
to $\gamma_2=\alpha+1$, and hence
\begin{equation}
\Delta\gamma=q-1.
\end{equation}
We apply these relations to the retained posterior samples of the proton-only 2CPL+BPL chain, obtaining $q=1.89^{+0.16}_{-0.11}$ and $\alpha=2.44^{+0.14}_{-0.09}$.
Repeating the transformation for the joint chain checks how these conditional transport parameters change in the joint analysis.
The joint posterior gives $(q_p,\alpha_p)=(1.75^{+0.08}_{-0.06},2.24^{+0.05}_{-0.04})$ and $(q_{\rm He},\alpha_{\rm He})=(1.80^{+0.08}_{-0.07},2.34^{+0.07}_{-0.06})$.
The species-dependent values are effective estimates from independently fitted spectral slopes, not evidence that protons and helium experience different turbulence.
These relations assume steady transport and an unchanged injection spectrum across the break; contributions from source-population averaging would alter their interpretation.
The break rigidity marks a balance between diffusion and drift escape rates, which depend on both transport coefficients and Galactic field geometry.
It therefore constrains combinations of field strength, turbulence and spatial scales, rather than uniquely measuring a coherence length by equating it to the particle gyroradius.

\end{document}